\documentclass[conference,a4paper]{IEEEtran}

\usepackage{amsmath,amssymb}
\usepackage{url}
\usepackage{graphicx}
\usepackage{bm}

\newtheorem{definition}{Definition}

\newtheorem{lemma}{Lemma}

\newcommand{\qed}{\hfill $\square$ \par}

\newcommand{\tr}{\mathrm{tr}}
\newcommand{\rmd}{\mathrm{d}}
\newcommand{\rmi}{\mathrm{i}}

\newcommand{\llangle}{\langle\kern -.23em \langle}
\newcommand{\rrangle}{\rangle\kern -.23em \rangle}

\renewcommand{\vec}[1]{\boldsymbol{#1}}

\begin{document}

\title{
  Generating Functional Analysis \\
  of Iterative Sparse Signal Recovery Algorithms \\ 
  with Divergence-Free Estimators
}
\author{
  \IEEEauthorblockN{Kazushi Mimura}
  \IEEEauthorblockA{
    Hiroshima City University \\
    3-4-1 Ohtsuka-higashi, Asaminami-ku, Hiroshima 731-3194, Japan \\
    Email: mimura@hiroshima-cu.ac.jp
  }
}
\maketitle
\begin{abstract}
Approximate message passing (AMP) is an effective iterative sparse recovery algorithm for linear system models. 
Its performance is characterized by the state evolution (SE) which is a simple scalar recursion. 
However, depending on a measurement matrix ensemble, AMP may face a convergence problem. 
To avoid this problem, orthogonal AMP (OAMP), which uses de-correlation linear estimation 
and divergence-free non-linear estimation, was proposed by Ma and Ping. 
They also provide the SE analysis for OAMP. 
In their SE analysis, the following two assumptions were made: 
(i) The estimated vector of the de-correlation linear estimator consists of 
i.i.d. zero-mean Gaussian entries independent of the vector to be estimated 
and (ii) the estimated vector of the divergence-free non-linear estimator consists of 
i.i.d. entries independent of the measurement matrix and the noise vector. 
In this paper, 
we derive a simple scalar recursion to characterize 
iterative sparse recovery algorithms with divergence-free estimators 
without such assumptions of independence of messages by using the generating functional analysis (GFA), 
which allows us to study the dynamics by an exact way in the large system limit. 
\end{abstract}



\section{Introduction}
\par
We consider a sparse signal recovery problem that 
a vector $\vec{x}\in\mathbb{R}^N$ is estimated from 
a measurement vector $\vec{y}\in\mathbb{R}^M \, (M < N)$: 
\begin{equation}
  \vec{y} = A \vec{x}_0 + \vec{\omega}, 
\end{equation}
where 
$A\in\mathbb{R}^{M \times N}$ denotes a measurement matrix 
and $\vec{\omega} \in \mathbb{R}^M$ denotes a noise vector $\vec{\omega} \sim \mathcal{N}(\vec{0},\sigma_0^2 I)$ 
\cite{Claerbout1973, Donoho1989, Candes2005, Donoho2006, Candes2006, Candes2006b}. 
The ratio $\delta = M/N$ is called the {\it compression rate}. 
When $\delta<1$, the system of equations undetermined. 
We assume the following to simplify the problem. 
Each entry of the {\it original vector} to be estimated $\vec{x}_0 =(x_{0,n}) \in \mathbb{R}^N$, is 
an i.i.d. random variable which obeys the distribution $p_{X_0}$, 
e.g., the Bernoulli-Gaussian distribution $p_{X_0}(x_0)=(1-\epsilon) \delta(x_0) + \epsilon (2\pi)^{-1/2} \exp(-x_0^2/2)$, 
where $\delta(x)$ denotes Dirac's delta function. 
The ratio between the number of non-zero entries and the dimension of the original vector is called the {\it signal density}. 
\par
Approximate message passing (AMP) is an effective iterative sparse recovery algorithm for linear system models \cite{Donoho2009}. 
Its performance is characterized by the state evolution (SE) which is a simple scalar recursion \cite{Donoho2009}. 
However, it is known that AMP may face a convergence problem, depending on a measurement matrix ensemble 
\cite{Rangan2014, Caltagirone2014, Rangan2016, Rangan2017, Takeuchi2017}. 
To avoid this problem and 
to improve performance, 
orthogonal AMP (OAMP), which uses de-correlation linear estimation 
and divergence-free non-linear estimation, was proposed by Ma and Ping \cite{Ma2017}. 
They also provide the SE analysis for OAMP \cite{Ma2017} under the following two assumptions: 
(i) The estimated vector of the de-correlation linear estimator consists of 
i.i.d. zero-mean Gaussian entries independent of the vector to be estimated 
and (ii) the estimated vector of the divergence-free non-linear estimator consists of 
i.i.d. entries independent of the measurement matrix and the noise vector. 
Bayati and Montanari have provided the rigorous foundation to SE for AMP \cite{Bayati2010}. 
On the other hand, SE for OAMP still needs theoretical justification. 
\par
In \cite{Mimura2011}, we have shown 
the {\it iterative shrinkage-thresholding algorithm} (IST) \cite{Donoho2009, Zibulevsky2010}, 
which cannot cancel the correlation between the present messages and their past values 
by applying the generating functional analysis (GFA) 
\cite{DeDominicis1978, Coolen2000, Heimel2001, Coolen2005, Mimura2014} that can treat complex correlation. 
In GFA, we assume that the generating functional is concentrated around its average over the randomness in the large system limit, 
and we use the saddle-point methods to calculate the generating functional asymptotically. 
\par
In this paper, 
we derive a simple scalar recursion to characterize 
iterative sparse recovery algorithms with divergence-free estimators 
without previous assumptions of independence of messages by using GFA. 
Although the advantage of OAMP is 
to be able to treat general unitarily-invariant matrices and various linear estimators,  
we here restricted ourselves to 
i.i.d. Gaussian matrices as measurement matrices and 
the matched filter as a linear estimator for simplicity of the analysis. 
\par
This paper is organized as follows. 
The next section introduces notations and algorithms. 
Section III explains the analysis. 
Section IV is for discussion. 
The final section is devoted to a summary.

\section{Preliminaries}
\par
Boldface lowercase letters and boldface uppercase letters denote vectors and matrices, respectively. 
We use the following notations: 
$\vec{0}$ for a vector and a matrix with all-zero entries, 
$\vec{1}$ for the identity matrix, 
$\vec{A}^\top$ for the transpose of $\vec{A}$, 
$\mathbb{E}_X$ for the expectation operation over a random variable $X$, and 
$\mathcal{N}(\vec{\mu},\vec{\Sigma})$ for the Gaussian distribution with mean vector $\vec{\mu}$ and covariance matrix $\vec{\Sigma}$. 

\subsection{De-Correlated and Divergence-Free Properties}
\par
These concepts are introduced by Ma and Ping \cite{Ma2017}. 
\par
\begin{definition}[De-Correlated Property (Definition 1, \cite{Ma2017})]
  For a given matrix $\vec{A}$, 
  if $\tr (\vec{1} - \vec{W}\vec{A}) = 0$ holds, 
  we say the matrix $\vec{W}$ is {\it de-correlated}. 
  \qed
\end{definition}
\par
When $\vec{A}$ is a matrix whose entries are i.i.d. Gaussian with mean zero 
and variance $1/M$, $\vec{A}^\top$ is a de-correlated against $\vec{A}$. 

\par
\begin{definition}[Divergence-Free Property (Definition 2, \cite{Ma2017})]
  For any $\tau \ge 0$ and any distribution $p_{X_0}$, if 
  \begin{align}
    \mathbb{E}_{x_0, z}[\eta'(x_0 + \tau z)]=0.
  \end{align}
  holds, we say the function $\eta : \mathbb{R} \to \mathbb{R}$ is {\it divergence-free}, 
  where 
  $x_0$ denotes a random variables $x_0 \sim p_{x_0}$ and 
  $z \sim \mathcal{N}(0,1)$ is a standard Gaussian random variable that is independent of $x_0$. 
  Here, $\eta'$ denotes a derivative of $\eta$. 
  \qed
\end{definition}

\subsection{AMP}
\par
Donoho et al. have proposed the following iterative algorithm achieving the performance of LP-based reconstruction. 
\begin{definition} \label{def:AMP}
  Starting from an initial guess $\vec{x}^{(0)}=\vec{0}$ and $\vec{z}^{(0)}=\vec{y}$, 
  approximate message passing (AMP) iteratively proceeds by 
  \begin{align}
    \vec{x}^{(t+1)} =& \; \eta_t(\vec{A}^\top \vec{z}^{(t)} + \vec{x}^{(t)}), \\
    \vec{z}^{(t)} =
    &\; \vec{y} - \vec{A} \vec{x}^{(t)} \nonumber \\
    &\; + \frac 1{\delta} \vec{z}^{(t-1)} \langle \eta_{t-1}' (\vec{A}^\top \vec{z}^{(t-1)} 
        + \vec{x}^{(t-1)}) \rangle. \label{eq:LE_AMP}
  \end{align}
  Here, $\{\eta_t\}$ is an appropriate sequence of threshold functions (applied componentwise for a vector), 
  $\vec{x}^{(t)} \in \mathbb{R}^N$ is the current estimate of the original vector $\vec{x}_0$, 
  $A^\top$ denotes the transpose of $A$ 
  and $\eta_t'(u)=\partial \eta_t(u) / \partial u$. 
  For a vector $\vec{v}=(v_1, \cdots, v_N)$, $\langle \vec{v} \rangle := N^{-1} \sum_{n=1}^N v_n$. 
  \qed
\end{definition}
While the property of AMP is investigated theoretically and thoroughly \cite{Donoho2009, Bayati2010},

\subsection{Orthogonal AMP}
\par
Ma and Ping have proposed OAMP as follows. 
\begin{definition} \label{def:OAMP}
  Starting from an initial guess $\vec{x}^{(0)}=\vec{0}$, 
  orthogonal approximate message passing (OAMP) iteratively proceeds by 
  \begin{eqnarray}
    \vec{x}^{(t+1)} &=& \eta_t( \textstyle \vec{W}^{(t)} \vec{z}^{(t)} + \vec{x}^{(t)}), \label{eq:OAMP} \\
    \vec{z}^{(t)} &=& \vec{y} - \vec{A} \vec{x}^{(t)} \label{eq:OAMP2}
  \end{eqnarray}
  where $\vec{W}^{(t)}$ is a de-correlated matrix 
  and $\eta_t$ is a divergence-free function for any $t$. 
  \qed
\end{definition}
In OAMP, the Onsager term, which is the third term in right hand side of (\ref{eq:LE_AMP}), vanishes. 
OAMP can treat general unitarily-invariant matrices $\vec{A}$ and various linear estimators $\vec{W}^{(t)}$.

\section{Analysis}
\par
We here restricted ourselves to 
i.i.d. Gaussian matrices as measurement matrices $\vec{A}$ and 
the matched filter as a linear estimator, i.e., $\vec{W}^{(t)}=\vec{A}^\top$ for all $t$, for simplicity of the analysis. 
Each entry of the {\it compression matrix} $\vec{A}=(a_{mn}) \in \mathbb{R}^{M \times N}$ is 
an i.i.d. Gaussian random variable of mean zero and variance $M^{-1}$, i.e., $a_{mn} \sim \mathcal{N}(0,M^{-1})$. 
We use GFA to derive a simple scalar recursion to characterize 
iterative sparse recovery algorithms with divergence-free estimators. 
So, we first introduce GFA and then start derivation.

\subsection{Genarating Functional Analysis}
\par
We analyze the dynamics in the large system limit where $N,M\to\infty$, while the compression rate $\delta=M/N$ is kept finite. 
The dynamics described by (\ref{eq:OAMP}) and (\ref{eq:OAMP2}) is a simple Markov chain, so the path probability $p[\vec{x}^{(0)},\cdots,\vec{x}^{(t)}]$, 
which is referred to as {\it path probability}, 
are given by products of the individual transition probabilities of the chain: 
\begin{align}
  p[\vec{x}^{(0)},\cdots,\vec{x}^{(t)}] 
  =& \delta [ \vec{x}^{(0)} ] \prod_{s=0}^{t-1} \delta [ \vec{x}^{(s+1)} \nonumber \\
   & - \eta_t ( \textstyle A^\top (\vec{y} - A \vec{x}^{(t)}) + \vec{x}^{(t)} + \vec{\theta}^{(t)}) ]. 
  \label{eq:updating_rule}
\end{align}
Here, $\vec{\theta}^{(t)}$ is an external message which is introduced to evaluate the response function 
and these parameters $\{ \vec{\theta}^{0}, \cdots, \vec{\theta}^{(t)} \}$ are set to be zero in the end of analysis. 
The initial state probability becomes $p[\vec{x}^{(0)}]=\prod_{n=1}^N \delta[x_n^{(0)}]$. 
Therefore, we can calculate an expectation with respect to an arbitrary function 
${\cal G}={\cal G}(\vec{x}^{(0)},\cdots,\vec{x}^{(t)})$ of tentative decisions as 
$\mathbb{E}_{\vec{x}}({\cal G}) \triangleq$ 
$\int_{\mathbb{R}^{(t+1)N}} ( \prod_{s=0}^t d\vec{x}^{(s)} )$ 
$p[\vec{x}^{(0)},\cdots,\vec{x}^{(t)}] {\cal G}$, 
where $\vec{x}$ denotes a set $\{\vec{x}^{(0)},\cdots,\vec{x}^{(t)}\}$ 
We define the following functional that is called the {\it generating functional} to analyze the dynamics of the system. 
\begin{definition}
  The generating functional $Z[\vec{\psi}]$ is defined by 
  \begin{align}
    Z[\vec{\psi}] \triangleq 
    \mathbb{E}_{\vec{x}} \biggl( 
      \exp \biggl[ - \rmi \sum_{s=0}^t\vec{x}^{(s)}\cdot\vec{\psi}^{(s)} \biggr] 
    \biggr) ,
    \label{eq:def_Z}
  \end{align}
  where $\vec{\psi}^{(s)}=$ $(\psi_1^{(s)},$ $\cdots,$ $\psi_N^{(s)})^\top$. 
  \qed
\end{definition}
In familiar way \cite{DeDominicis1978, Coolen2000, Mimura2014}, 
one can obtain all averages of interest by differentiation, e.g., 
\begin{align}
  i\lim_{\vec{\psi}\to\vec{0}}\frac{\partial Z[\vec{\psi}]}{\partial \psi_n^{(s)}}
  &= \mathbb{E}_{\vec{x}} (x_n^{(s)}), \\ 
  -\lim_{\vec{\psi}\to\vec{0}}\frac{\partial Z[\vec{\psi}]}{\partial \psi_n^{(s)} \partial \psi_{n'}^{(s')}} 
  &= \mathbb{E}_{\vec{x}} (x_n^{(s)} x_{n'}^{(s')} ), \\ 
  i\lim_{\vec{\psi}\to\vec{0}}\frac{\partial Z[\vec{\psi}]}{\partial \psi_n^{(s)} \partial \theta_{n'}^{(s')}} 
  &= \frac{\partial \mathbb{E}_{\vec{x}} ( x_n^{(s)} ) }{\partial \theta_{n'}^{(s')}}.  
\end{align}
from $Z[\vec{\psi}]$. 
We assume that the generating functional is concentrated to its average 
over the random variables $\{A, \vec{x}_0, \vec{\omega} \}$ in the large system limit, 
namely the typical behavior of the system depends only on the statistical properties of the random variables. 
We therefore evaluate the averaged generating functional 
$\bar{Z}[\vec{\psi}] = \mathbb{E}_{\vec{x}, A, \vec{x}_0, \vec{\omega}} 
( \exp [ - \rmi \sum_{s=0}^t\vec{x}^{(s)}\cdot\vec{\psi}^{(s)} ] ) $, 
where $\overline{[\cdots]}$ denotes an expectation over $\{A, \vec{x}_0, \vec{\omega} \}$. 
Evaluating the averaged generating functional, one can obtain important parameters that describe the algorithm performance. 
Namely, we can evaluate the overlap, which is also called the direction cosine, 
between the original vector $\vec{x}_0$ and the current estimate $\vec{x}^{(s)}$ 
and the second moment of the current estimate. 
Since $||\vec{x}_0 - \vec{x}^{(t)}||_2^2 = ||\vec{x}_0||_2^2 - 2\vec{x}^{(t)} \cdot \vec{x}_0 + ||\vec{x}^{(t)}||_2^2$, 
we can evaluate MSE from the overlap and the second moment. 
Here, $\vec{x}^{(t)} \cdot \vec{x}_0$ denotes the inner product between $\vec{x}^{(t)}$ and $\vec{x}_0$.

\subsection{Outline of the Analysis}
\par
We apply GFA to the algorithm written by (\ref{eq:OAMP}) and (\ref{eq:OAMP2}). 
One can obtain the following result. 
\begin{lemma} \label{lemma:gfa}
  For IST with an arbitrary sequence of threshold functions $\{\eta_s\}_{s=0}^t$, 
  MSE per component $\sigma_t^2$ of the current estimate $\vec{x}^{(t)}$ can be assessed as 
  \begin{align}
    \sigma_t^2 
    &:= N^{-1} \mathbb{E}_{\vec{x}, A, \vec{x}_0, \vec{\omega}}(||\vec{x}_0 - \vec{x}^{(t)}||_2^2) \notag\\
    &= \mathbb{E}_{x_0}[(x_0)^2]-2m^{(t)}+C^{(t,t)} 
    \label{eq:MSE}
  \end{align}
  in the large system limit, i.e., $N\to\infty$, where the parameters are given as follows. 
  \begin{align}
    m^{(s)}    &= \llangle x_{0} x^{(s)} \rrangle, \label{eq:sp_m} \\
    C^{(s,s')} &= \llangle x^{(s)} x^{(s')} \rrangle, \\
    G^{(s,s')} &= \frac{\partial \llangle x^{(s)}\rrangle}{\partial \theta^{(s')}} \mathbb{I}(s>s'), 
  \end{align}
  where $\mathbb{I}(\mathcal{P})$ denotes an indicator function which takes 1 
  if the proposition $\mathcal{P}$ is true, 0 otherwise. 
  Here, the average over the effective path measure $\llangle \cdots \rrangle$ is given by 
  \begin{align}
    & \llangle g(\vec{x},\vec{v}) \rrangle \notag\\
    & :=  \mathbb{E}_{x_0} \biggl( \int_{\mathbb{R}^t} \mathcal{D}\vec{v} \int_{\mathbb{R}^{t+1}} \biggl( \prod_{s=0}^t \! dx^{(s)} \biggr) g(\vec{x},\vec{v}) \; \delta[x^{(0)}] \notag \\
    & \times \prod_{s=0}^{t-1} \! \delta \bigl[ x^{(s+1)} - \eta_s\bigl(x_0 \hat{k}^{(s)}+v^{(s)}+(\vec{\Gamma} \vec{x})^{(s)} + \theta^{(s)} \bigr) \bigr] \biggr), \label{eq:Dv}
  \end{align}
  where 
  \begin{align}
    & {\cal D}\vec{v} = |2\pi \vec{R}|^{-1/2} d\vec{v} \exp [ -\frac 12\vec{v}\cdot \vec{R}^{-1}\vec{v}], \\ 
    & \vec{R} = (\vec{1}+\delta^{-1} \vec{G}^\top)^{-1} \vec{D} (\vec{1}+\delta^{-1} \vec{G})^{-1}, \\ 
    & \vec{\Gamma} = (\vec{1}+\delta^{-1} \vec{G})^{-1} \delta^{-1} \vec{G}, \\
    & \hat{k}^{(s)} = |\vec{\Lambda}_{[s]}|. 
  \end{align} 
  Each entries of $\vec{D}$ and $\vec{\Lambda}_{[s]}$ are 
  \begin{align}
    & D^{(s,s')} = \sigma_0^2 + \delta^{-1} ( \mathbb{E}_{x_0}[(x_0)^2] -m^{(s)}-m^{(s')}+C^{(s,s')} ], \\
    & \Lambda_s^{(s',s'')} = \delta_{s,s'}+(1-\delta_{s,s'})(\delta_{s',s''} + \delta^{-1} G^{(s'',s')}), 
  \end{align}
  respectively. 
  The terms $(\vec{R}^{-1}\vec{v})^{(s)}$ and $(\vec{\Gamma} \vec{\sigma})^{(s)}$ denote 
  the $s^{\rm th}$ entry of the vector $\vec{R}^{-1}\vec{v}$ and $\vec{\Gamma} \vec{\sigma}$, respectively. 
  We put $\theta^{(0)}=\cdots=\theta^{(t)}=0$. 
  \qed
\end{lemma}
\par
Brief outline of derivation is given in Appendix A, which is almost same to the analysis of \cite{Mimura2011}. 
\par
In GFA, we extract a one-dimensional iterative process which is statistically equivalent to the original $N$-dimensional iterative process. 
The effective path measure $\llangle \cdots \rrangle$ is an expectation operator with respect to such a one-dimensional process. 
Lemma \ref{lemma:gfa} entirely describe the dynamics of the system. 
The term $(\vec{\Gamma} \vec{\sigma})^{(s)}$ in (\ref{eq:Dv}) corresponds to the Onsager term. 
\par
We next derive the following Lemma. 
\begin{lemma}
  \label{lemma:G=0}
  For any $t \ge 0$, $\vec{G}=\vec{0}$ holds, if $\eta_0, \cdots, \eta_t$ are divergence-free. 
  \qed 
\end{lemma}
To show this lemma, we use the inductive method. 
\par
(i) The case of $t=0$. 
Since this is an initial step, the Onsager term does not exist. 
We have $G^{(0,0)}=0$ in this case. 
The response function matrix becomes $\vec{G}=\vec{0} \in \mathbb{R}^{1 \times 1}$ (scalar). 
We then have $\vec{\Gamma}=\vec{0} \in \mathbb{R}^{1\times 1}$ (scalar). 
We also have 
$\vec{\Lambda}_{[0]}=\vec{1} \in \mathbb{R}^{1\times 1}$ (scalar), 
$\hat{k}^{(0)}=\det \vec{\Lambda}_{[0]} =1$, 
and $\vec{R} \in \mathbb{R}^{1 \times 1}$ (scalar) as Lemma \ref{lemma:gfa}.

\par
(ii) The case of $t=1$. 
The response function $G^{(1,0)}$ is given as 
\begin{align}
  G^{(1,0)} 
  =& \frac{\partial \mathbb{E}_{x_0} [ \llangle x^{(1)} \rrangle ]}{\partial \theta^{(0)}} \biggr|_{\vec{\theta}=\vec{0}} \notag\\
  =& \frac{\partial}{\partial \theta^{(0)}} 
  \mathbb{E}_{x_0} \biggl( \int_{\mathbb{R}} {\cal D}\vec{v} \int_{\mathbb{R}^{2}} dx^{(0)} dx^{(1)} \, x^{(1)} \, \delta [x^{(0)}] \notag \\
  & \times \delta \bigl[ x^{(s+1)} - \eta_0 \bigl(x_0 \hat{k}^{(s)}+v^{(s)}+(\vec{\Gamma} \vec{x})^{(s)} + \theta^{(s)} \bigr) \bigr] \biggr) \biggr|_{\vec{\theta}=\vec{0}}
  \notag\\
  =& \frac{\partial}{\partial \theta^{(0)}} 
  \mathbb{E}_{x_0} \int_{\mathbb{R}} \frac{dv^{(0)} e^{-\frac12v^{(0)} (R^{(0,0)})^{-1} v^{(0)}}}{\sqrt{|2\pi R^{(0,0)}|}} \notag\\
  &\times \eta_0 (x_0 +v^{(0)} + \theta^{(0)} ) |_{\vec{\theta}=\vec{0}} \notag\\
  =& \mathbb{E}_{x_0, z} [ \eta_0' (x_0 + \sqrt{R^{(0,0)}} z ) ] \notag\\
  =& 0, 
\end{align}
where $z \sim \mathcal{N}(0,1)$ is independent of $x_0 \sim p_{x_0}(x_0)$. 
Note that we use the divergence-free property in the last equality. 
Since $\mathbb{I}(1>1)=\mathbb{I}(0>1)=0$, we immediately have $G^{(1,1)}=G^{(0,1)}=0$. 
We then have $\vec{G}=\vec{0} \in \mathbb{R}^{2 \times 2}$. 
We also have 
$\vec{\Gamma}=(\vec{1}+\frac1\delta\vec{G})^{-1}\delta\vec{G}=\vec{0} \in \mathbb{R}^{2 \times 2}$ and 
\begin{align}
  \vec{\Lambda}_{[1]}=\left(
  \begin{array}{cc}
    1 & 0 \\
    1 & 1 \\
  \end{array}
  \right).
\end{align}
We therefore obtain $\hat{k}^{(1)}=\det \vec{\Lambda}_{[1]} =1$.

\par
(iii) The case of $t=2$. 
The expectation value of $x^{(2)}$ is given as 
\begin{align}
  &\!\!\!\!
  \mathbb{E}_{x_0} [ \llangle x^{(2)} \rrangle ] \notag\\
  =& 
  \mathbb{E}_{x_0} \biggl( \int_{\mathbb{R}^2} {\cal D}\vec{v} \int_{\mathbb{R}^{3}} dx^{(0)} dx^{(1)} dx^{(2)} \, x^{(2)} \, \delta [x^{(0)}] \notag \\
  & \times \delta \bigl[ x^{(1)} - \eta_0\bigl(x_0 \hat{k}^{(0)}+v^{(0)}+(\vec{\Gamma} \vec{x})^{(0)} + \theta^{(0)} \bigr) \bigr] \notag\\
  & \times \delta \bigl[ x^{(2)} - \eta_1\bigl(x_0 \hat{k}^{(1)}+v^{(1)}+(\vec{\Gamma} \vec{x})^{(1)} + \theta^{(1)} \bigr) \bigr] \biggr)
  \notag\\
  =& \mathbb{E}_{x_0, z} [ \eta_1' (z_0 + \sqrt{R^{(1,1)}} z + \theta^{(1)} ) ]. 
\end{align}
Since the expectation value of $x^{(2)}$ does not contain $\theta^{(0)}$, the response function $G^{(2,0)}$ is 
\begin{align}
  G^{(2,0)} 
  = \biggl( \frac{\partial}{\partial \theta^{(0)}} 
  \mathbb{E}_{x_0, z} [ \eta_1 (x_0 + \sqrt{R^{(1,1)}} z + \theta^{(1)} ) ] \biggr) \biggr|_{\vec{\theta}=\vec{0}} 
  \!\!\!\!
  = 0. 
\end{align}
On the other hand, the response function $G^{(2,1)}$ also becomes 
\begin{align}
  G^{(2,1)} 
  = \mathbb{E}_{x_0, z} [ \eta_1' (x_0 + \sqrt{R^{(1,1)}} z ) ] 
  = 0. 
\end{align}
by using divergence-free property. 
Since $\mathbb{I}(2>2)=\mathbb{I}(1>2)=\mathbb{I}(0>2)=0$, we immediately have $G^{(2,2)}=G^{(1,2)}=G^{(0,2)}=0$. 
We then have $\vec{G}=\vec{0} \in \mathbb{R}^{3 \times 3}$. 
We also have 
$\vec{\Gamma}=(\vec{1}+\frac1\delta\vec{G})^{-1}\delta\vec{G}=\vec{0} \in \mathbb{R}^{3 \times 3}$ and 
\begin{align}
  \vec{\Lambda}_{[2]}=\left(
  \begin{array}{ccc}
    1 & 0 & 0 \\
    0 & 1 & 0 \\
    1 & 1 & 1 \\
  \end{array}
  \right).
\end{align}
In this case, we also have $\hat{k}^{(2)}=\det \vec{\Lambda}_{[2]} =1$.

\par
(iv) The case of $t>2$. 
We next show that $\vec{G}=\vec{0} \in \mathbb{R}^{(t+1) \times (t+1)}$ by using the inductive method. 
In the case of $t=s$, we here assume that $\vec{G}=\vec{0} \in \mathbb{R}^{(s+1) \times (s+1)}$. 
Under this assumption, we have $\vec{\Gamma}=\vec{0}$ and $\hat{k}^{(0)}= \cdots \hat{k}^{(s)}=1$. 
The expectation value of $x^{(s+1)}$ is
\begin{align}
  & \!\!\!\! \mathbb{E}_{x_0} [ \llangle x^{(s+1)} \rrangle ] \notag\\
  =& \mathbb{E}_{x_0} \biggl( \int_{\mathbb{R}^{s+1}} {\cal D}\vec{v} \int_{\mathbb{R}^{s+2}} dx^{(0)} \cdots dx^{(s+1)} \, x^{(s+1)} \notag \\
  & \times \delta [x^{(0)}] \notag \\
  & \times \delta \bigl[ x^{(1)} - \eta_0\bigl(x_0 \hat{k}^{(0)}+v^{(0)}+(\vec{\Gamma} \vec{x})^{(0)} + \theta^{(0)} \bigr) \bigr] \notag\\
  & \times \cdots \notag\\[-2mm]
  & \times \delta \bigl[ x^{(s+1)} - \eta_s \bigl(x_0 \hat{k}^{(s)}+v^{(s)} 
  +(\vec{\Gamma} \vec{x})^{(s)} + \theta^{(s)} \bigr) \bigr] \biggr) \notag\\
  =& 
  \mathbb{E}_{x_0} \int_{\mathbb{R}} \frac{dv^{(s)} e^{-\frac12v^{(s)} (R^{(s,s)})^{-1} v^{(s)}}}{\sqrt{|2\pi R^{(s,s)}|}} 
  \eta_{s} (x_0 +v^{(s)} + \theta^{(s)} ) \notag\\
  =& \mathbb{E}_{x_0, z} [ \eta_s' (x_0 + \sqrt{R^{(s,s)}} z + \theta^{(s)} ) ]. 
\end{align}
Since $\mathbb{E}_{x_0} [ \llangle x^{(s+1)} \rrangle ]$ contains $\theta^{(s)}$ only, 
for $s' \in \{0, \cdots, s-1\}$, the response function $G^{(s,s')}$ becomes 
\begin{align}
  G^{(s,s')} 
  = \biggl( \frac{\partial}{\partial \theta^{(s')}} 
  \mathbb{E}_{x_0, z} [ \eta_s (x_0 + \sqrt{R^{(s,s)}} z + \theta^{(s)} ) ] \biggr) \biggr|_{\vec{\theta}=\vec{0}} 
  \!\!\!\!
  = 0. 
\end{align} 
Using divergence-free property, we have 
\begin{align}
  G^{(s,s-1)} 
  = \mathbb{E}_{x_0, z} [ \eta_s' (x_0 + \sqrt{R^{(s,s)}} z ) ]
  = 0. 
\end{align}
Since we have $G^{(s+1,s+1)} =G^{(s,s+1)} = \cdots =G^{(0,s+1)}=0$, 
we obtain $\vec{G}=\vec{0} \in \mathbb{R}^{(s+2) \times (s+2)}$. 
\par
If the claim holds for $t=s$, it holds for $t=s+1$. 
This proves Lemma \ref{lemma:G=0}. 
\par
We next evaluate MSE. 
We can write a closed from equation for MSE by using only the diagonal entries of the covariance matrix $\vec{R}$. 
We define the effective noise variance $\tau_t^2$ as 
\begin{align}
  \tau_t^2 := R^{(t,t)}. 
\end{align}
Using $\vec{G}=\vec{0}$, the covariance matrix becomes
\begin{align}
  \vec{R} 
  = (\vec{1}+\delta^{-1} \vec{G}^\top)^{-1} \vec{D} (\vec{1}+\delta^{-1} \vec{G})^{-1} 
  = \vec{D} 
\end{align}
whose $(t,t')$ entry is given as 
\begin{align}
  D^{(t,t')}
  = \sigma_0^2 + \delta^{-1} (\mathbb{E}_{x_0}[(x_0)^2]-2m^{(t)}+C^{(t,t')}). 
\end{align}
We then have 
\begin{align}
  & \tau_t^2 = \sigma_0^2 + \frac1\delta \sigma_t^2, \\
  & \sigma_{t+1}^2 = \mathbb{E}_{x_0,z} [\{x_0-\eta_t(x_0+\tau_t z)\}^2].
\end{align}
This coincides with the result of SE for OAMP obtained by assuming independence of messages.

\section{Discussion}
\par
In \cite{Ma2017}, to apply SE, the OAMP error recursion is first introduced as follows: 
\begin{align}
  & \vec{h}^{(t)} = (\vec{1}-\vec{W}^{(t)}\vec{A}) \vec{q}^{(t)} + \vec{W}^{(t)} \vec{n}, \\
  & \vec{q}^{(t+1)} = \eta_t(\vec{x}_0 + \vec{h}^{(t)}) - \vec{x}_0. 
\end{align}
Next, the following two assumptions were made: 
(i) $\vec{h}^{(t)}$ consists of i.i.d. zero-mean Gaussian entries independent of $\vec{x}_0$ for every $t$, and 
(ii) $\vec{q}^{(t+1)}$ consists of i.i.d. entries independent of $\vec{A}$ and $\vec{n}$. 
GFA reveals that these two assumptions are not required to derive SE 
for iteration algorithms with divergence-free estimators. 
Our result means that just the divergence-free property is sufficient to cancel the complex correlation of past values correctly.

\section{Summary}
\par
We derive a simple scalar recursion to characterize 
iterative sparse recovery algorithms with divergence-free estimators. 
By applying GFA, we show that the assumptions of messages are not required to derive the SE recursion equation. 
which allows us to study the dynamics by an exact way in the large system limit. 
This result gives theoretical justification of SE for OAMP derived in \cite{Ma2017}. 
The analysis in the case of more general measurement matrices, such as unitarily-invariant matrices, and linear estimator is now underway.

\section*{Acknowledgment}
\par
The author would like to thank Junjie Ma, Arian Maleki, Keigo Takeuchi, and Jun'ichi Takeuchi for their fruitful discussions. 
This work was partially supported by 
JSPS Grant-in-Aid for Challenging Exploratory Research Grant Number 16K12496, 
JSPS Grant-in-Aid for Scientific Research (B) Grant Number 16H02878, 
and JST CREST Grant Number JPMJCR15G1, Japan.

\appendix
\subsection{Outline of analysis of Lemma 1}

\par
Let $\vec{u}^{(t)}=(u_n^{(t)})$ be a summation of messages, i.e., 
$\vec{u}^{(t)} \triangleq A^\top \vec{z}^{(t)} + \vec{x}^{(t)} + \vec{\theta}^{(t)}$, 
where $\vec{\theta}^{(t)}$ is an external message which is introduced to evaluate the response function $G^{(s,s')}$. 
The Dirac's delta function is replaced as $\delta(x) = \gamma (2\pi)^{-1/2} e^{-\gamma^2 x^2/2}$ and 
the parameter $\gamma$ is taken the limit $\gamma \to \infty$ later. 
We first separate the summation of messages at any iteration step by inserting the following delta-distributions: 
$ 1 = \int \delta\vec{u}\delta\hat{\vec{u}} \prod_{s=0}^{t-1} \prod_{n=1}^N \exp [ i\hat{u}_n^{(s)} \{ u_n^{(s)} - ( A^\top \vec{z}^{(t)} )_n - x_n^{(t)} - \theta_n^{(t)} \} ]$, 
where $\delta\vec{u} \triangleq \prod_{s=0}^{t-1} \prod_{n=1}^N \frac{du_n^{(s)}}{\sqrt{2\pi}}$ 
and $\delta\hat{\vec{u}} \triangleq \prod_{s=0}^{t-1} \prod_{n=1}^N \frac{d\hat{u}_n^{(s)}}{\sqrt{2\pi}}$. 
Here, $( \vec{a} )_n$ denotes the $n^{\mathrm{th}}$ element of the vector $\vec{a}$. 
\par
The disorder-averaged generating functional is for $N\to\infty$ dominated by a saddle-point \cite{Copson1965, Merhav2009}. 
We can thus simplify the saddle-point problem to (\ref{eq:Z3}). 
The disorder-averaged generating functional is then simplified to the saddle-point problem as 
\begin{align}
  \bar{Z}[\vec{\psi}] 
  =& \mathbb{E}_{\vec{x}_0} \biggl( \int 
  \rmd\vec{\mathsf{m}} d\hat{\vec{\mathsf{m}}}
  \rmd\vec{k} d\hat{\vec{k}}
  \rmd\vec{q} d\hat{\vec{q}}
  \rmd\vec{Q} d\hat{\vec{Q}}
  \rmd\vec{L} d\hat{\vec{L}} \nonumber \\
  & \times \exp \biggl[ N(\Psi+\Phi+\Omega)+O(\ln N) \biggr] \biggr),   \label{eq:Z3}
\end{align}
in which the functions $\Psi$, $\Phi$, $\Omega$ are given by 
\begin{align}
  \Psi \triangleq
  & i \sum_{s=0}^{t-1} \{ \hat{\mathsf{m}}^{(s)} \mathsf{m}^{(s)} + \hat{k}^{(s)}k^{(s)} \} + i \sum_{s=0}^{t-1} \sum_{s'=0}^{t-1}  \{ \hat{q}^{(s,s')} q^{(s,s')} \nonumber \\
  & + \hat{Q}^{(s,s')} Q^{(s,s')} + \hat{L}^{(s,s')} L^{(s,s')} \} 
\end{align}
\begin{align}
  \Phi \triangleq
  & \frac 1N \sum_{n=1}^N \ln \biggl\{ \int_{\mathbb{R}^{t+1}} \biggl( \prod_{s=0}^{t-1} x^{(s)} \biggr) p[x^{(0)}] \int \delta u \delta\hat{u} \nonumber \\
  & \times \exp \biggl[ \sum_{s=0}^{t-1} \{ \ln \frac{\gamma}{\sqrt{2\pi}} - \frac{\gamma^2}2 [x^{(s+1)}-\eta_s(u^{(s)})]^2 \} \nonumber \\
  & - i \sum_{s=0}^{t-1} \sum_{s'=0}^{t-1} \{ \hat{q}^{(s,s')} x^{(s)} x^{(s')} \nonumber \\
  & \quad + \hat{Q}^{(s,s')} \hat{u}^{(s)} \hat{u}^{(s')} + \hat{L}^{(s,s')} x^{(s)} \hat{u}^{(s')} \} \nonumber \\
  & + i \sum_{s=0}^{t-1} \hat{u}^{(s)} \{ u^{(s)} - x^{(s)} - \theta_n^{(s)} - x_{0,n} \hat{k}^{(s)} \} \nonumber \\
  & - i \sum_{s=0}^{t-1} x_{0,n} x^{(s)} \hat{\mathsf{m}}^{(s)} - i \sum_{s=0}^t x^{(s)} \psi_n^{(s)} \biggr] 
\end{align}
\begin{align}
  \Omega \triangleq
  & \frac 1N \ln \int \delta\vec{v}\delta\hat{\vec{v}}\delta\vec{w}\delta\hat{\vec{w}} \nonumber \\
  & \times \exp \biggl[ i \sum_{m=1}^M \sum_{s=0}^{t-1} \{ \hat{v}_m^{(s)} v_m^{(s)} + \hat{w}_m^{(s)} w_m^{(s)} - \frac1{\delta} v_m^{(s)} w_m^{(s)} \} \nonumber \\
  & - \frac 12 \sum_{m=1}^M \sum_{s=0}^{t-1} \sum_{s'=0}^{t-1} \{ \frac1{\delta} \sigma_0^2 v_m^{(s)} v_m^{(s')} + \hat{v}_m^{(s)} Q^{(s,s')} \hat{v}_m^{(s')} \} \nonumber \\
  & - \frac 12 \sum_{m=1}^M \sum_{s=0}^{t-1} \sum_{s'=0}^{t-1} \{ \hat{v}_m^{(s)} [k^{(s)} - L(s',s)] \hat{w}_m^{(s')} \nonumber \\
  & \quad + \hat{w}_m^{(s)} [k^{(s')} - L^{(s,s')}] \hat{v}_m^{(s')} \} \nonumber \\
  & - \frac 12 \sum_{m=1}^M \sum_{s=0}^{t-1} \sum_{s'=0}^{t-1} \! \{ \! \hat{w}_m^{(s)} [ \mathbb{E}_{x_0}[(x_0)^2] - \! \mathsf{m}^{(s)} \! - \! \mathsf{m}^{(s')} \! + \! q^{(s,s')}] 
  \hat{w}_m^{(s')} \} \biggr]
\end{align}
where $\delta u \triangleq \prod_{s=0}^{t-1} \frac{du^{(s)}}{\sqrt{2\pi}}$ and $\delta \hat{u} \triangleq \prod_{s=0}^{t-1} \frac{d\hat{u}^{(s)}}{\sqrt{2\pi}}$. 
In the limit $N\to\infty$, the integral (\ref{eq:Z3}) will be dominated by the saddle point of the extensive exponent $\Psi+\Phi+\Omega$. 
Straightforward differentiation and taking the limit $\gamma \to \infty$, 
we then arrive at Lemma \ref{lemma:gfa}.



\begin{thebibliography}{99}

\bibitem{Claerbout1973}
J.F.Claerbout and F.Muir, 
``Robust modeling with erratic data,'' 
{\it Geophysics}, vol.38, no.5, pp.826--844, Oct.1973. 

\bibitem{Donoho1989}
D.L.Donoho, 
``Uncertainty Principles and Signal Recovery,'' 
{\it SIAM J. Appl. Math.}, vol.49, no.3, pp.906--931, 1989. 

\bibitem{Candes2005}
E.J.Cand\'es and T.Tao, 
``Decoding by linear programming,'' 
{\it IEEE Trans. Info. Theory}, vol.51, no.12, pp.4203--4215, Dec.2005. 

\bibitem{Donoho2006}
D.L.Donoho, 
``Compressed Sensing,'' 
{\it IEEE Trans. Info. Theory}, vol.52, no.4, pp.1289--1306, Apr.2006. 

\bibitem{Candes2006}
E.J.Cand\'es, J.Romberg, and T.Tao, 
``Robust uncertainty principles: exact signal reconstruction from highly incomplete frequency information,'' 
{\it IEEE Trans. Info. Theory}, vol.52, no.2, pp.489--509, Feb.2006.  

\bibitem{Candes2006b}
E.J.Cand\'es and T.Tao, 
``Near-Optimal Signal Recovery From Random Projections: Universal Encoding Strategies?'' 
{\it IEEE Trans. Info. Theory}, vol.52, no.12, pp.5406--5425, Dec.2006. 

\bibitem{Donoho2009}
D.L.Donoho, A.Maleki and A.Montanari, 
``Message-passing algorithms for compressed sensing,'' 
{\it Proc. of the National Academy of Sciences (PNAS)}, vol.106, no.45, pp.18914--18919, Sep.2009. 







\bibitem{Rangan2014}
S.Rangan, P.Schniter, and A.Fletcher, 
``On the convergence of Approximate Message Passing with Arbitrary Matrices,'' 
{\it Proc. of ISIT2014}, pp.236--240, Jul.2014. 

\bibitem{Caltagirone2014}
F.Caltagirone, L.Zdeborov\'a, and Florent Krzakala
``On Convergence of Approximate Message Passing,'' 
{\it Proc. of ISIT2014}, pp.1812--1816, Jul.2014. 

\bibitem{Rangan2016}
S.Ranga, 
``Fixed Points of Generalized Approximate Message Passing With Arbitrary Matrices,'' 
{\it IEEE Trans. Info. Theory}, vol.62, no.12, pp.7464--7474, Dec.2016. 

\bibitem{Rangan2017}
S.Rangan, P.Schniter, and A.K.Fletcher, 
``Vector approximate message passing,'' 
{\it Proc. of ISIT2014}, pp.1588--1592, Jul.2017. 

\bibitem{Takeuchi2017}
K.Takeuchi, 
``Rigorous Dynamics of Expectation-Propagation-Based Signal Recovery from Unitarily Invariant Measurements,'' 
{\it Proc. of ISIT2014}, pp.501--505, Jul.2017. 





\bibitem{Ma2017}
J. Ma and L. Ping, 
``Orthogonal AMP,'' 
{\it IEEE Access}, vol.5, pp.2020--2033, Jan.2017. 

\bibitem{Bayati2010}
M.Bayati and A.Montanari, 
``The Dynamics of Message Passing on Dense Graphs, with Applications to Compressed Sensing,'' 
{\it IEEE Trans. Info. Theory}, vol.57, no.2, pp.764--785, Feb.2011. 

\bibitem{Mimura2011}
K. Mimura, 
``Generating functional analysis of iterative reconstruction algorithms for compressed sensing,'' 
{\it Proc. of ISIT2009}, pp.1432--1436, Aug.2011. 




\bibitem{Zibulevsky2010}
M. Zibulevsky and M. Elad, 
``L1-L2 Optimization in Signal and Image Processing,'' 
{\it IEEE Signal Proc. Mag.}, vol.27, no.3, pp.76--88, Apr.2010. 







\bibitem{DeDominicis1978}
C.De Dominicis, 
``Dynamics as a substitute for replicas in systems with quenched random impurities,'' 
{\it Phys. Rev. B}, vol.18, 4913, Nov.1978. 

\bibitem{Coolen2000}
A.C.C.Coolen, 
``Statistical mechanics of recurrent neural networks II. Dynamics,'' 
{\it Preprint} arXiv cont-mat/0006011, Jun.2000. 

\bibitem{Heimel2001}
J.A.F.Heimel, 
``Dynamics of learning by neurons and agents,'' 
{\it Doctoral Thesis}, King's College London, 2001. 

\bibitem{Coolen2005}
A.C.C.Coolen, 
{\it The Mathematical Theory of Minority Games}, 
Oxford Univ. Press, 2005.

\bibitem{Mimura2014}
K.Mimura and M.Okada, 
``Generating Functional Analysis for Iterative CDMA Multiuser Detectors,'' 
{\it IEEE Trans. Info. Theory}, vol.60, no.6, pp.3645-3670, Jun.2014. 














\bibitem{Copson1965}
E.T.Copson, 
{\it Asymptotic Expansions}, 
Cambridge Univ. Press, 1965. 

\bibitem{Merhav2009}
N.Merhav, 
``Statistical Physics and Information Theory,'' 
now publishing, 2009. 












\end{thebibliography}
\end{document}